# Enhancement of giant refrigerant capacity in $Ho_{1-x}Gd_xB_2$ alloys ($0.1 \leq x \leq 0.4$)

Pedro Baptista de Castro[1,2]*, Kensei Terashima[1]*, Takafumi D Yamamoto[1], Suguru Iwasaki[3], Akiko T. Saito[1], Ryo Matsumoto[1], Shintaro Adachi[4], Yoshito Saito[1,2], Mohammed ElMassalami[5], Hiroyuki Takeya[1], Yoshihiko Takano[1,2]

[1]*National Institute for Materials Science, 1-2-1 Sengen, Tsukuba, Ibaraki 305-0047, Japan*

[2]*University of Tsukuba, 1-1-1 Tennodai, Tsukuba, Ibaraki 305-8577, Japan*

[3]*Hokkaido University, Laboratory of Nanostructured Functional Materials, Research Institute for Electronic Science (RIES), N20 W10, Kita-ku, Sapporo, Hokkaido 001-0020, Japan*

[4]*Kyoto University of Advanced Science (KUAS), Ukyo-ku, Kyoto 615-8577, Japan*

[5]*Instituto de Fisica, Universidade Federal do Rio de Janeiro, CxP 68528, 21945-972 Rio de Janeiro, Brazil*

E-mail: CASTRO.Pedro@nims.go.jp, TERASHIMA.Kensei@nims.go.jp

**Abstract**

Intending to optimize the giant magnetocaloric properties of $HoB_2$, we synthesized and magnetocalorically characterized $Ho_{1-x}Gd_xB_2$($0.1 \leq x \leq 0.4$) alloys. We found out that Gd enters stoichiometrically and randomly into the Ho site, leading to a Vegard-type structural change. The addition of spherical $S^{7/2}$ $Gd^{3+}$ moments prompts an enhancement in Curie temperature, a reduction in peak value of the magnetic entropy change while still being relatively high, and a broadening of the magnetic entropy change curves. The overall influence is a relatively high refrigerant capacity and relative cooling power, and an extension of the thermal working range to higher temperatures; thus, electing $Ho_{1-x}Gd_xB_2$ as potential candidates for cryogenic refrigeration applications.

# 1. Introduction

The magnetocaloric effect (MCE) is often characterized by the magnitude of the induced change in magnetic entropy ($\Delta S_M$) when a magnetic field is varied. It proves to be useful for wide and diverse technological applications in areas ranging from medicine[1,2] to solid-state magnetic refrigerators[3–5]. Magnetic refrigerators, in particular, manifest a potential not only for replacing the traditional gas-based refrigerators (due to its higher efficiency, compactness, environmental friendliness, and noiselessness[4,6–8] but also for cryogenic applications such as gas liquefaction. Particular interest has been focused on the applicability of MCE to hydrogen[9,10] liquefication/refrigeration since such technology is expected to play a key role in the so-called Hydrogen society. For that purpose, one requires materials with excellent MCE character, preferably allowing them to be operated within the range, say, flanked by hydrogen and nitrogen liquefaction points:[9] $T_{LH}$= 20.3 K ~ $T_{LN}$= 77 K. A common practice for securing such MCE materials is to optimize the MCE response of suitable giant MCE materials by an adequate alloying, $x$, which leads to (i) changing $T_C(x)$ such that $T_{LH} < T_C(x) < T_{LN}$, (ii) expanding the useful thermal working range by broadening the $\Delta S_M$ curves, (iii) maintaining a relatively high refrigerant capacity, and (iv) avoiding any hysteresis effects.

Recently, we have reported the remarkable MCE properties of the $HoB_2$ compound[11] which is characterized as having a high $|\Delta S_M|$ = 40.1 J kg K$^{-1}$ (0.35 J cm$^{-3}$ K$^{-1}$) near its $T_C$ of 15 K, for $\mu_0 \Delta H$ = 5 T.

Below we briefly consider alloying-based optimization of MCE properties in this material. A partial substitution on the B sites of $HoB_2$ has already been attempted, and the investigation of Ho(B$_{1-x}$Si$_x$)$_2$ shown structural stability up to $x$ = 0.2 and that HoB$_{1.6}$Si$_{0.4}$ manifests AFM order with $T_N$ = 11 K[12]. Regarding the $R$-site option, the $R$B$_2$ structure in the stoichiometric compound is limited to Tb-Lu[13–15] and recalling that for $T_C$ of DyB$_2$ = 50-55 K[12,16] and TbB$_2$ = 151 K[14], the $T_C(x)$ in R-site of substituted alloys is expected to follow the effective de Gennes factor, as such scaling has been seen in other $R$-$R$ alloys[17], intermetallics $R$-$M$ [18], and $R$-nitrides[19] systems. The particular case of Ho$_{1-x}$Dy$_x$B$_2$ ( 0 ≤ $x$ ≤1) is promising since both $T_C(x)$ and the broadness of the $\Delta S_M$ curves are reasonably increased[20] but it happened that weak hysteresis effects are manifested. Thus, we opted for alloying with Gd, as we expect that the substitution of the spherical S$^{7/2}$ character of Gd$^{3+}$ moments (Ho$_{1-x}$Gd$_x$B$_2$), to satisfy the conditions (i)-(iv).

In this work, we report on our studies on the MCE of Ho$_{1-x}$Gd$_x$B$_2$ alloys (0≤$x$≤0.4): It is shown that incorporation of Gd$^{3+}$ moments brings about all the beforementioned advantages: most prominently, it enhances $T_C$ (doubled at $x$ = 0.4) and broadens the peak of |Δ$S_M$| curve leading to an enhancement of the refrigerant capacity and relative cooling power.

## 2. Experimental Methods

Polycrystalline samples of Ho$_{1-x}$Gd$_x$B$_2$ were synthesized via conventional argon arc-melting procedures using stoichiometric amounts of Ho (99.9%), Gd (99.9%), and B (99.5%). Samples were subjected to repeated intermediate grinding, pelletizing, and arc-melting procedures. We managed to obtain single-phase for $x$ = 0.2, 0.3, 0.4 samples; however, despite the various attempts, no single-phase samples with $x ≥ 0.5$ were obtained; rather multiple phases, namely Gd$_2$B$_5$ and $R$B$_4$ ($R$=Gd, Ho) were obtained, indicating that $x$ = 0.4 is close to the solubility limit. This is consistent with the phase diagram of Gd-B system which identifies the metastable character of the GdB$_2$ phase[21]. This phase diagram also explains the tendency of as-prepared Ho$_{1-x}$Gd$_x$B$_2$ samples ($x > 0.4$) to decompose during post-arc-melting heat treatment.

For the crystal structure analysis, powder X-ray diffractograms (XRD) of the obtained samples were measured using a MiniFlex 600 Rigaku diffractometer with Cu K$_\alpha$ radiation.

The magnetization was measured under zero-field cooling (ZFC) and field cooling (FC) protocol within an MPMS XL (Quantum Design) setup for the analysis of the magnetic properties of the samples. In the case of obtaining the magnetic entropy change, isofield ($M$-$T$) curves in a wide range of applied fields up to 5 T was taken using a ZFC protocol.

## 3. Results and Discussion

Figure 1 (a) shows the diffractograms of the studied samples. Except for $x$ = 0.1 (with weak contamination of $R_2$O$_3$ of less than 3%), all samples are single-phase AlB$_2$-type ($P6/mmm$) structure. This was confirmed by using Rietveld-refinement as implemented within the FULLPROF package[22]. The linear evolution of the lattice parameters, shown in Fig. 1 (b) and 1(c), follows Vegard`s law confirming the successful substitutional incorporation of Gd up to $x$=0.4.

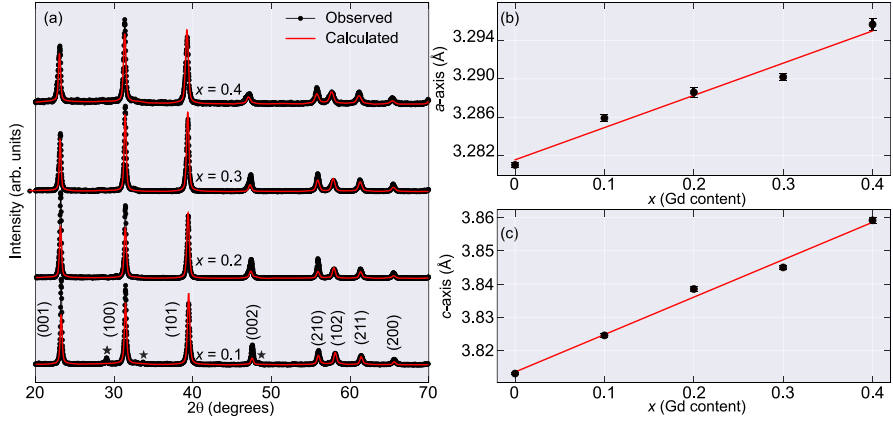

**Figure 1.** Powder XRD patterns and obtained lattice parameters. (a) Powder XRD patterns for Ho$_{1-x}$Gd$_x$B$_2$ alloys. The red lines show the calculated curves, (b) Evolution of the *a*-axis parameter, and (c) Evolution of the *c*-axis parameter as a function of Gd content. The red line is a linear fit emphasizing Vegard's law. The data for $x = 0$ was taken from Ref. [20]

Figure 2 (a) shows the *M-T* magnetization curves of Ho$_{1-x}$Gd$_x$B$_2$ under $\mu_0 H = 0.01$ T. The evolution of $M(x, T, \mu_0 H = 0.01$ T$)$ can best be understood if we recall that HoB$_2$ is a ferromagnet with $T_C \approx 15$ K which, below $T^* \approx 11$ K, is possibly undergoing a spin-reorientation process[11]. Accordingly, $M(x, T, \mu_0 H = 0.01$T$)$ of Ho$_{1-x}$Gd$_x$B$_2$ can be envisaged as paramagnetic (with reversible and hysteresis-less ZFC and FC curves) above $T_C(x)$, ferromagnetic below $T_C(x)$ with a possible spin-reorientation at $T^* \approx 11$ K. The evolution of each of $T_C(x)$ and $T^*(x)$ can be identified from the ZFC $\partial M/\partial T$ $(x, T, \mu_0 H = 0.01$ T$)$ curves shown in Figure 2(b). Evidently, $T_C(x)$ is appreciably enhanced on increasing $x$ (15, 19, 22, 27, and 30 K, see Fig 4. (a)), most probably related to the increase in the effective de Gennes factor. A similar increase in $T_C(x)$ was reported in Ho$_{1-x}$Dy$_x$B$_2$[20]. In contrast $T^*(x) \approx 13$ K, is very weakly influenced by $x$, this might be understood in such a way that the incorporation of the spherically $S^{7/2}$-state hardly influences the angular reorientation.

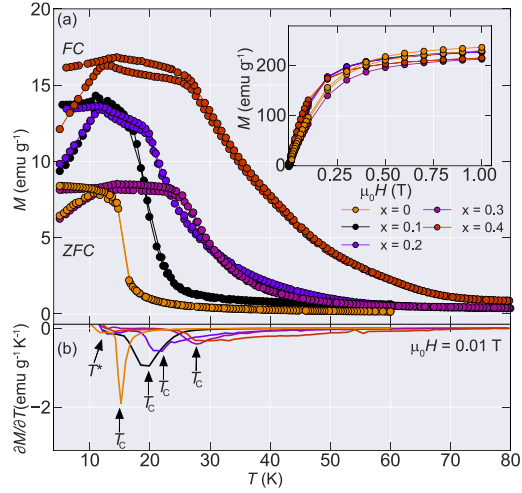

**Figure 2.** Isofield and isothermal magnetization of $Ho_{1-x}Gd_xB_2$ alloys. (a) ZFC and FC isofield $M(x, T, \mu_0H = 0.01\ T)$. The inset shows the isothermal $M(x, T = 5\ K, \mu_0H)$. (b) The temperature-dependent derivatives of the ZFC curves. The ferromagnetic transitions and the lower temperature transition at $T = 13$ K, are marked by arrows. The data for $x = 0$ was taken from Ref. [20], and its derivative value is divided by a factor of 2 for clarity.

The isothermal $M(x, T = 5\ K, \mu_0H)$ curves, shown in the inset of Figure 2(a), manifests an easy saturation, under relatively weak applied fields. Also, there is no strong hysteresis effects: a desirable feature when using $Ho_{1-x}Gd_xB_2$ in cryogenic refrigeration applications.

The influence of Gd incorporation on the MCE of $Ho_{1-x}Gd_xB_2$ was evaluated based on extensive $M(x, T, \mu_0H)$ curves, over a wide range of fields from 0.01 T to 5 T. These are shown in Figs. 3 (a)-(d). The corresponding $|\Delta S_M(x, T, \mu_0\Delta H)|$ curves were evaluated via the Maxwell relation:

$$\Delta S_M = \mu_0 \int_0^H \left(\frac{\partial M}{\partial T}\right) dH. \qquad [1]$$

The obtained $|\Delta S_M(x, T, \mu_0\Delta H)|$ curves are shown in Figs. 3 (e)-(h). Here, volumetric units (J cm$^{-3}$K$^{-1}$) are mainly used, instead of the gravimetric units (J kg$^{-1}$ K$^{-1}$), since these are the most convenient unit for application purposes[23,24].

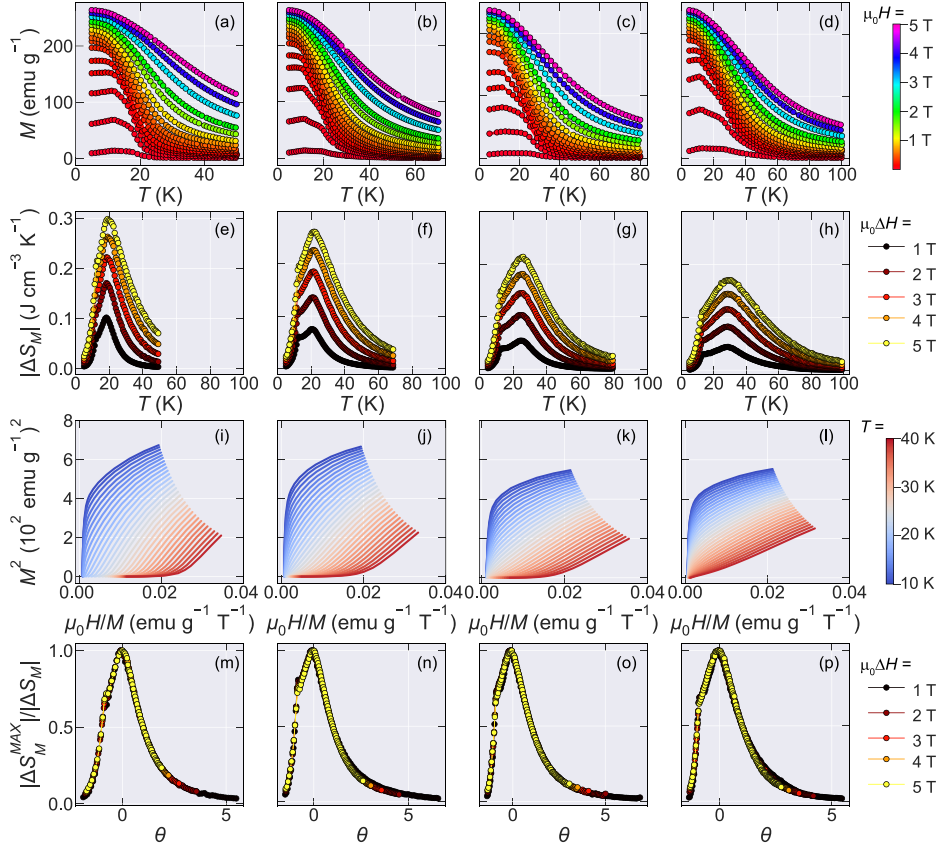

**Figure 3** ZFC isofield magnetization , calculated $|\Delta S_M(x, T, \mu_0\Delta H)|$, Arrott, and normalized entropy curves. (a)-(d) ZFC $M(x,T, \mu_0 H)$ curves measured under various ranges $0.1 \leq x \leq 0.4$, $0.01 \leq \mu_0 H \leq 5$ T, and $2 \leq T \leq 100$K. (e-h) Calculated $|\Delta S_M(x, T, \mu_0\Delta H)|$ curves using Eq.1 and the magnetization curves of panels (a)-(d). (i-l) Arrott plots derivated from the isofield magnetization curves of panels (a-d). (m-p) Normalized entropy *versus* normalized temperatures. These are constructed from the curves shown in panels (e-h).

Three features are worthy of highlighting: (i) The maximum in entropy change, $|\Delta S_M^{MAX}|$ for $\mu_0\Delta H = 5$T is monotonically decreased (0.3, 0.26, 0.21, 0.17 J cm$^{-3}$ K$^{-1}$ for $x = 0.1, 0.2, 0.3, 0.4$, respectively, see Fig 4(b) and Table 1). (ii) The point marking $|\Delta S_M^{MAX}(x)|$ coincides with $T_c(x)$: both are monotonically increased. (iii) The monotonic variations in $|\Delta S_M^{MAX}(x)|$ and $T_c(x)$ are accompanied by a broadening in the width of the MCE peak (see Figs. 2(b) and 4 (c)) which is usually defined as the temperature span $\delta T_{FWHM}(x, \mu_0\Delta H) = T_{r_1} - T_{r_2}$. $T_{r_1}$ and $T_{r_2}$ are defined as the points at which $|\Delta S_M(x, T_{r_i}, \mu_0\Delta H)| = 0.5|\Delta S_M^{MAX}(x, \mu_0\Delta H)|$. Part of the increase in $\delta T_{FWHM}$ may be due to the presence of the second transition at $T^*$(which is

also evident as a low-temperature peak in the entropy curves). This increase has a significant impact on the refrigerant capacity (*RC*) and relative cooling power (*RCP*) defined as:[4,23–25]

$$RC(x, \Delta\mu_o H) = \int_{T_{r_1}}^{T_{r_2}} \Delta S_M(x, T, \mu_o \Delta H) dT \quad [2]$$

$$RCP(x, \mu_o \Delta H) = \Delta S_M^{MAX}(x, \mu_o \Delta H) * \delta T_{FWHM}(x, \mu_o \Delta H) \quad [3]$$

For $\mu_0\Delta H$ = 5T, one obtains *RC* (*RCP*) = 5.61 (7.17), 6.07 (7.68), 5.64 (7.09) and 5.94 (7.48) J cm$^{-3}$ for $x$ = 0.1, 0.2, 0.3 and 0.4 respectively (see Figs. 4 (d)-(e)). Evidently, as $x$ is increased, the $|\Delta S_M(x, T, \mu_0\Delta H)|$ curve approach the desirable tablelike MCE feature without a strong variation in the *RC* (*RCP*) parameter. We summarize the magnetocaloric properties of these alloys in Table 1, where we show both volumetric and gravimetric units.

| Alloy | $\Delta S$ | | RC | | RCP | |
|---|---|---|---|---|---|---|
| | J cm$^{-3}$ K$^{-1}$ | J kg$^{-1}$ K$^{-1}$ | J cm$^{-3}$ | J kg$^{-1}$ | J cm$^{-3}$ | J kg$^{-1}$ |
| Ho$_{0.9}$Gd$_{0.1}$B$_2$ | 0.3 | 34.6 | 5.61 | 651 | 7.17 | 833 |
| Ho$_{0.8}$Gd$_{0.2}$B$_2$ | 0.26 | 30.6 | 6.07 | 711 | 7.68 | 899 |
| Ho$_{0.7}$Gd$_{0.3}$B$_2$ | 0.21 | 25.2 | 5.64 | 665 | 7.09 | 836 |
| Ho$_{0.6}$Gd$_{0.4}$B$_2$ | 0.17 | 20.2 | 5.94 | 707 | 7.48 | 889 |

**Table 1** The magnetocaloric and figure of merit properties of the obtained Ho$_{1-x}$Gd$_x$B$_2$ alloys for $\mu_0\Delta H$ = 5 T.

On comparing Ho$_{1-x}$Gd$_x$B$_2$ with the end material HoB$_2$, one expects the magnetic phase transitions at $T_C$ in Ho$_{1-x}$Gd$_x$B$_2$ alloys to be second-order. Indeed, this is confirmed from the Arrott plots of Fig. 2 (i)-(l) (taken within the $10 \leq T \leq 40$ K range at steps of 1 K): the slope of each curve is positive emphasizing, based on Banerjee criterion[26], the second-order character. An alternative demonstration of the second-order character is usually given in terms of the universal scaling curve[27–29] depicting the normalized entropy change, $|\Delta S_M(x, T, \mu_0\Delta H)| / |\Delta S_M^{MAX}(x, T = T_C, \mu_0\Delta H)|$, as a function of the normalized temperature $\theta(x, T, \mu_0\Delta H)|$, defined as:

$$\theta(x, T, \mu_0 \Delta H) = \begin{cases} -(T - T_c)/(T_{r_1} - T_c), & T \leq T_c \\ (T - T_c)/(T_{r_2} - T_c), & T > T_c \end{cases}$$

where $T_{r_1}$ and $T_{r_2}$ are defined above. Recall that $T_C$, $T_{r_1}$ and $T_{r_2}$ are a function of $x$ and $\mu_0 \Delta H$. As evident, the normalized curves, Figure 3 (m)-(p), collapse on each other as expected for a second-order phase transition.

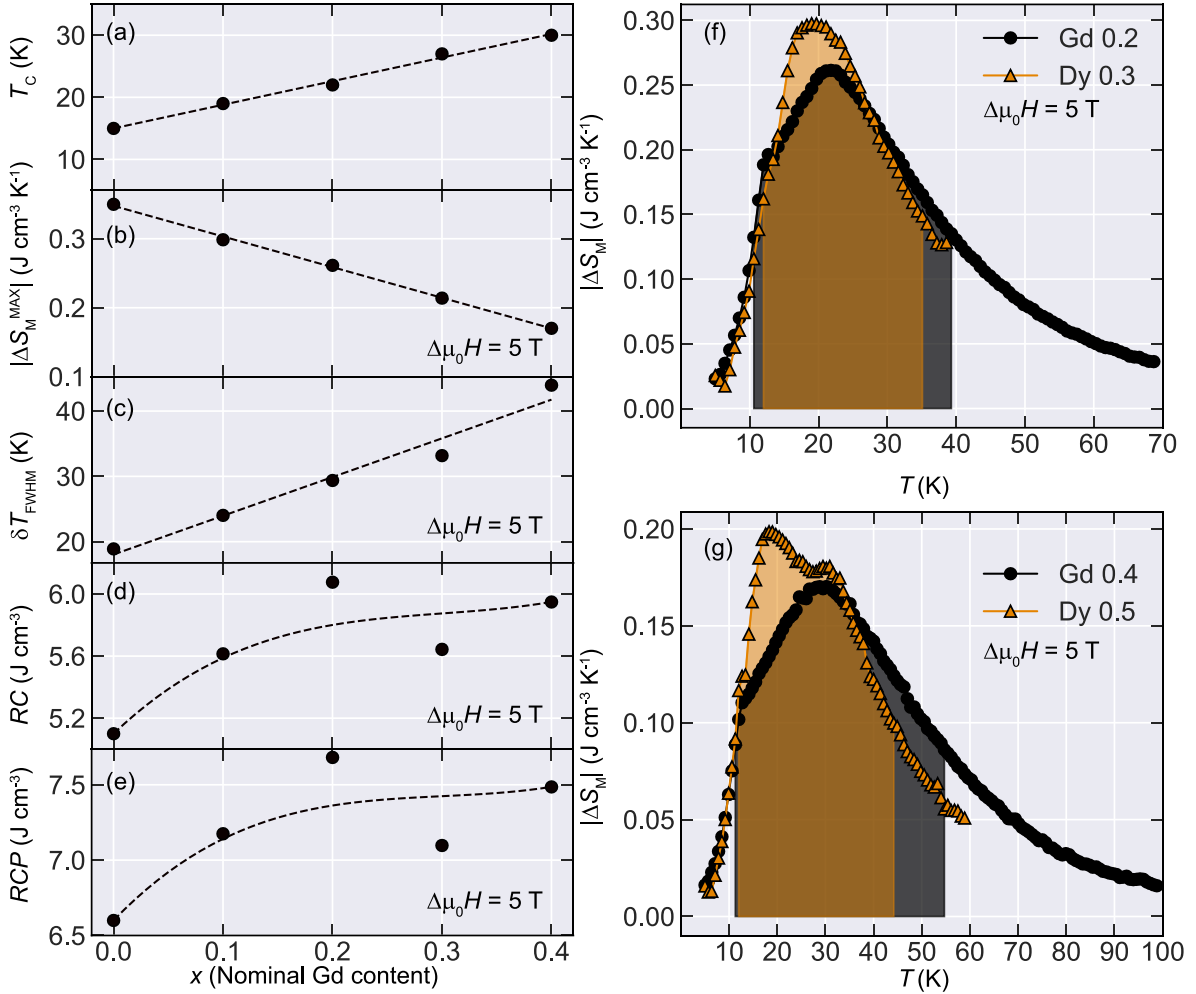

**Figure 4.** Evolution of the ferromagnetic transition $T_C$, maximum entropy change $|\Delta S_M^{MAX}|$, $\delta T_{FWHM}$, $RC$ and $RCP$ for $\mu_0 \Delta H = 5$ T. (a) $T_C$, (b) $|\Delta S_M^{MAX}|$, (c) $\delta T_{FWHM}$, (d) $RC$, and (e) $RCP$ versus $x$. The dashed lines are guides for the eyes. The data for $x = 0$ and Dy-substituted samples were taken from Ref. [20]. Comparison of $|\Delta S_M|$ (f) between $Ho_{0.8}Gd_{0.2}B_2$ and $Ho_{0.7}Dy_{0.3}B_2$, and (g) between $Ho_{0.6}Gd_{0.4}B_2$ and $Ho_{0.5}Dy_{0.5}B_2$. The painted areas show graphically the refrigerant capacity of each sample. The brown region indicates the overlapping region between the samples.

Finally, let us compare the magnetic and MCE properties of $Ho_{1-x}Gd_xB_2$ and $Ho_{1-x}Dy_xB_2$[20] (the only $Ho_{1-x}R_xB_2$ alloys with reported MCE properties). Three main differences can be identified: (i) In contrast to Dy ($0 \leq x \leq 1$), the solubility of Gd is limited to below $x=0.4$; (ii) the anisotropic character of $Dy^{3+}$ is relatively stronger than that of $Gd^{3+}$; (iii) The effective de Gennes factor of $Ho_{1-x}Gd_xB_2$ is higher than that of $Ho_{1-x}Dy_xB_2$. Let us also compare, the MCE properties for samples with similar $T_C$ in both series, namely $Ho_{0.8}Gd_{0.2}B_2$ with $Ho_{0.7}Dy_{0.3}B_2$ (both $T_C s \sim 22$ K, Fig.4 (f)) and $Ho_{0.6}Gd_{0.4}B_2$ with $Ho_{0.5}Dy_{0.5}B_2$ (both $T_C s \sim 30$ K, Fig.4 (g)). For both pairs, $Ho_{1-x}Gd_xB_2$ samples tend to exhibit almost comparable but a slightly lower $|\Delta S_M^{MAX} (T = T_C, \mu_0 \Delta H = 5$ T$)|$ than the $Ho_{1-x}Dy_xB_2$ samples by 0.1-0.4 J cm$^{-3}$ K$^{-1}$, while more significantly, *RC* and *RCP* of Gd containing samples are higher by a factor of 7-10% for RC and 7-15% for RCP compared to the Dy case, due to the higher broadness of the $\Delta S_M$ curves (compare the painted areas of Figs. 4 (f)-(g)). Given these facts, combined with the non-inducing hysteresis characteristic of Gd alloying, makes Gd alloying also a promising candidate for a temperature working range of $T < 30$ K.

4. **Conclusions**

In summary, we synthesized and systematically studied the magnetocaloric properties of $Ho_{1-x}Gd_xB_2$ alloys ($0.1 \leq x \leq 0.4$). We find out that incorporation of Gd leads to a Vegard-type structural change, an enhancement in $T_C$, a reduction in the maximum entropy change $|\Delta S_M^{MAX}|$, and a broadening of $\delta T_{FWHM}$. For all studied compositions, we observed a second-order magnetic phase transition at $T_C$ [evidence was obtained from Arrotts plots, universal scaling curves, and lack of appreciable hysteresis effects in $M(x, T = 5$ K, $\mu_0 H)$]. It is concluded that, within the $15 \leq T \leq 30$K range, these MCE features (namely the large volumetric entropy change, the broadened entropy curves, and the high *RC*/*RCP*) elect $Ho_{1-x}Gd_xB_2$ as potential candidates for magnetic-based hydrogen liquefaction/refrigeration applications. In fact, their MCE properties are comparable or even better than those reported for previous candidates such as $ErAl_2$ ($T_C = 14$ K)[30], HoN ($T_C = 18$ K[31]), $DyNi_2$ ($T_C = 21$ K)[30] and $HoAl_2$ ($T_C = 31.5$ K)[32] (For a detailed comparison with other MCE candidates, see Ref. 11).

**Acknowledgments**

This work was supported by the JST-Mirai Program (Grant No. JPMJMI18A3), JSPS KAKENHI (Grant Nos. 19H02177, 20K05070), JST CREST (Grant No. JPMJCR20Q4) P.B. Castro acknowledges the scholarship support from the Ministry of Education, Culture, Sports, Science and Technology (MEXT), Japan.